\documentclass[11pt,onecolumn,amssymb,nofootinbib]{revtex4}
\usepackage{bm}
\usepackage{amsmath}
\usepackage{amssymb}
\usepackage{graphicx}
\usepackage{amsmath}
\usepackage{bbm}

\begin{document}
\title{\bf On a recent proposal of faster than light quantum communication}

\author{Angelo Bassi}
\email{bassi@ts.infn.it}
\address{Dipartimento di Fisica Teorica,
Universit\`a di Trieste, Strada Costiera 11, 34014 Trieste, Italy.
\\  Mathematisches Institut der L.M.U., Theresienstr. 39, 80333
M\"unchen, Germany, \\
Istituto Nazionale di Fisica Nucleare, Sezione di Trieste, Strada
Costiera 11, 34014 Trieste, Italy.}

\author{GianCarlo Ghirardi}
\email{ghirardi@ts.infn.it}%
\affiliation{Dipartimento di Fisica Teorica dell'Universit\`a degli
Studi di Trieste, Strada Costiera 11, 34014 Trieste, Italy, \\
Istituto Nazionale di Fisica Nucleare, Sezione di Trieste, Strada
Costiera 11, 34014 Trieste, Italy, \\ The Abdus Salam International
Centre for Theoretical Physics, Strada Costiera 11, 34014 Trieste,
Italy}

\begin{abstract}
{\bf Abstract}. In a recent paper, A.Y. Shiekh has discussed an
experimental set-up which, in his opinion, should make possible
faster-than-light communication using the collapse of the quantum
wave function. Contrary to the many proposals which have been
presented in the past, he does not resort to an entangled state of
two systems but he works with a single particle in a superposition
of two states---corresponding to its propagation in opposite
directions---one of which goes through an appropriate
interferometer. The possibility  for an observer near the
interferometer to introduce or not, at his free will, a phase
shifter along one of the paths should allow to change
instantaneously the probability of finding the particle in the
far-away region corresponding to the other state of the
superposition and, correspondingly, to change the intensity of a
beam of particles reaching a distant observer. In this paper we show
a flaw in the argument: once more, as it has been proved in full
generality a long time ago, the process of wave packet reduction
cannot be used for superluminal communication.
\\

\end{abstract}
\maketitle

\section{Introduction}

In a recent paper~\cite{sh}, it has been suggested the possibility
of resorting to a specific experimental set up to send faster than
light signals by taking advantage of the process of wave packet
reduction. The aspect which makes the proposal completely different
from all those which have appeared in the literature up to now
derives from the fact that the author does not resort to an
entangled state of two systems one of which is subjected to a
measurement, but he works with a single particle in a superposition
of two states, and the measurement process as well as the ensuing
reduction of the wave packet (w.p.r.) involves only one of the two
far-away parts of the wave function. We recall that general proofs
that the quantum process of wave packet reduction does not allow
faster-than-light communication have been presented many years ago
\cite{eb,grw,gw,ggw}. However all the just mentioned proofs dealt
with  situations  involving two entangled states. So, in a sense,
the argument of Ref.~\cite{sh} does not fall under such general
proofs and requires a separate comment. In this brief paper we will
show that the process of w.p.r. cannot be used to send superluminal
signals to distant observers by following the procedure suggested by
Shiekh. Since the author has proposed~\cite{shbis} to resort to an
experimental setup identical to the one considered in Ref.~\cite{sh}
in order to enhance the efficiency of certain quantum information
protocols, our analysis proves the inapplicability also of these
proposals.

\section{Shiekh's argument}

We briefly review the argument by Shiekh~\cite{sh}. He considers a
particle which is prepared, at time $t=0$, in a equal weights
superposition\footnote{The fact that the two state have the same
coefficient, is irrelevant; both the argument of the author as well
as our counterargument hold for any superposition of the two
states.} of two normalized states, $\vert h+\rangle$ and $\vert
h-\rangle$, propagating in two opposite directions, respectively,
starting from the common origin of the $x$-axis:

\begin{equation} \label{eq:1}
\vert \psi,0\rangle=\frac{1}{\sqrt{2}}[\vert h+\rangle+\vert
h-\rangle].
\end{equation}

\noindent Subsequently  the state $\vert h+\rangle$ is injected in
an appropriate device behaving in a way similar, apart from the
final stage, to a Mach-Zender interferometer. This device, which
splits the term $|h+\rangle$ of the superposition in two components
propagating along two different paths, is positioned along the
positive real $x$-axis at an appreciable distance from the origin,
and an observer located near it can choose, at his free will, to
insert or not a phase-shifter along one of the two paths. The two
wave functions are then recombined by appropriate deflectors and it
is assumed that, by deciding whether or not to insert the
phase-shifter, one can produce a constructive (no phase-shifter in
place) or a destructive (the phase-shifter is present) interference
of the two terms in which the impinging state $\vert h+\rangle$ has
been split. Finally, a detector is placed along the direction of
propagation of the final state and it induces wave packet reduction,
since it either detects or fails to detect the particle. We have
summarized the  situation for the two considered cases in Figs.~1a
and~1b. The pictures simply reproduce those of Ref.~\cite{sh}.
\begin{figure}[t]
\begin{center}
\includegraphics[scale=0.6]{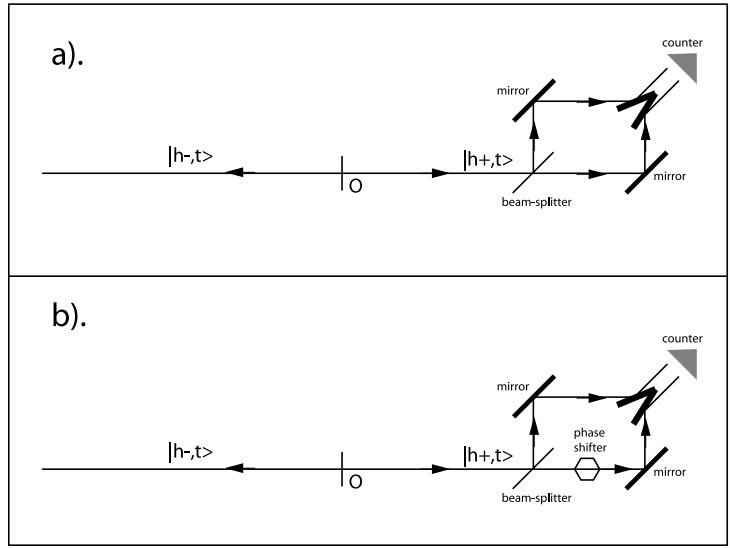}
\caption{The experimental arrangement considered in Ref.~\cite{sh}.
The two cases a) and b) correspond to no phase-shifter inserted or
the phase-shifter inserted in one of the arms of the device at
right, respectively.} \label{f1}
\end{center}
\end{figure}

The author then concludes: {\it If the sender} (a term used to
denote the man who, by his choice of inserting or not the phase-shifter, can change - according to the author - the  amplitude of
the far away wave function) {\it arranges for constructive
interference, then some of the particles will be "taken up" by the
sender, but none if destructive interference is arranged; in this
way the sender can control the intensity of the beam detected by the
receiver} (the observer located far away where the evolved of $\vert
h-\rangle$ is concentrated). {\it So, a faster than light
transmitter of information (but not energy or matter) might be
possible}. We now will reconsider this argument.

\section{The evolution of a wave function crossing the interferometric device}

In order to make clear the flaw in the argument of Ref.~\cite{sh},
let us focus our attention on a spatially well-localized and normalized wave packet
$\vert \phi \rangle$ propagating along the positive direction of the
$x$-axis, with an almost definite momentum, which is then injected
in the Mach-Zender like apparatus, as depicted in Fig.~2. For the
purpose of our analysis, $\vert \phi \rangle$ can represent one of
the following two situations:
\begin{itemize}
\item The term $| h+ \rangle$ of an initial superposition of two
wave packets, one traveling towards the interferometer, the other
traveling in the opposite direction, as in Shiekh's setup;
\item A {\it classical} light impulse, in such a way that
$|\langle x|\phi \rangle|^2$ is proportional to its energy density. This
possibility makes it clear that, as far as the behavior of
$|\phi\rangle$ within the interferometer (and prior to the
measurement) is concerned, nothing subtle, highly non-classical goes
on.
\end{itemize}
In the following, for the sake of simplicity, we will refer to
$|\phi\rangle$ as representing a quantum particle.

\begin{figure}[t]
\begin{center}
\includegraphics[scale=0.8]{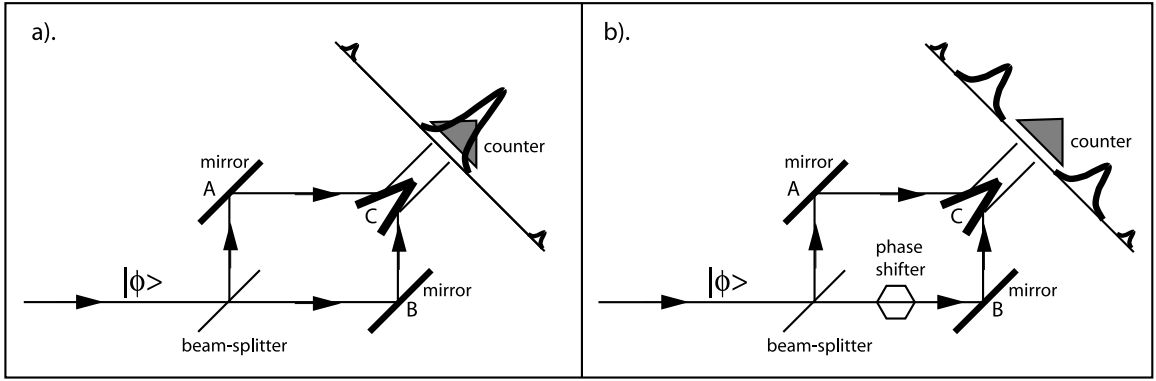}
\caption{The actual situation concerning the position probability
density distribution when constructive a) and destructive b) interference occurs.}
\label{f2}
\end{center}
\end{figure}

When the wave function $|\phi\rangle$ hits the first beam splitter,
half of it keeps moving horizontally and half
vertically\footnote{Here and in the following, we will ignore the
(possible) phase shifts introduced by the beam splitter and/or the
mirrors.}; the two mirrors $A$ and $B$ simply change the direction
of propagation of the two terms of the superposition; finally, when
meeting the mirrors $C$, each of the two terms of the superposition
is deflected, describing a particle at an almost definite position
propagating at $45^{0}$ with respect to the $x$-axis towards the
detector. In this way these two terms can give rise to an
interference pattern.

If we denote by $|\phi_{\ell}\rangle$ the part of the wave function
which followed the lower route in the interferometer, and by
$|\phi_{u}\rangle$ the part which followed the upper route, the
effect of the interferometer on the initial state $|\phi\rangle$ can
be summarized as follows:
\begin{equation} \label{eq:fin}
|\phi\rangle \; \rightarrow \; |\phi_{f}\rangle = \frac{1}{\sqrt{2}}[|\phi_{u}\rangle
\, + \, e^{i \varphi} |\phi_{\ell}\rangle],
\end{equation}
where $\varphi$ is the phase introduced by the phase shifter, i.e.
\begin{equation}
\varphi \; = \; \left\{
\begin{array}{ll}
0 \;\; & \text{phase shifter not inserted} \\
\pi & \text{phase shifter inserted.}
\end{array}
\right.
\end{equation}
\noindent Note that,  for the very structure of the  interferometer,
the statevectors $\vert\phi_{u}\rangle$ and
$\vert\phi_{\ell}\rangle$ represent two identical, well localized
nearby states propagating towards the detector.

Here comes the crucial point: the two mirrors in $C$ cannot be
designed in such a way to perfectly superimpose the two terms
$|\phi_{\ell}\rangle$ and $|\phi_{u}\rangle$ one right on top of the
other, in such a way to make them cancel each other when $\varphi =
\pi$. The reason is simply that such a device would violate
unitarity\footnote{If we think of $|\phi\rangle$ as representing a
classical beam of light propagating in space, then a similar device
would violate conservation of energy!}. A device behaving in this
way would be analogous to a glass which is perfectly transparent on
one side and perfectly reflecting on the other side: as shown in
Fig.3, it would allow, by an appropriate choiche of the two paths
and the insertion of a phase shifter on one of them, to get
perfectly destructive interference and, consequently, for quantum
particles (or classical beams of light) to disappear, which is not
possible and goes also against common sense. Accordingly, and as
depicted in Fig. 2 a) and b), the state $|\phi_{f}\rangle$ of
Eq.~\eqref{eq:fin} represents the superposition of two (essentially
identical) wave packets {\it traveling parallel to each other, at
most contiguous}, and such that $\langle \phi_{f} |\phi_{f}\rangle =
\langle \phi | \phi \rangle=1$, because of unitarity\footnote{Note
that the argument we have developed forbids, precisely for the same
reasons, also a partial attenuation of the incoming wave function,
i.e. that $\langle \phi_{f} |\phi_{f}\rangle <1$. Any decrease of
the probability of the particle being detected in a given interval
of the axis orthogonal to the propagation direction of the states
emerging from the interferometer must be accompanied by a
corresponding increase of finding the particle in the complement  of
the considered interval.}.

\begin{figure}[t]
\begin{center}
\includegraphics[scale=0.9]{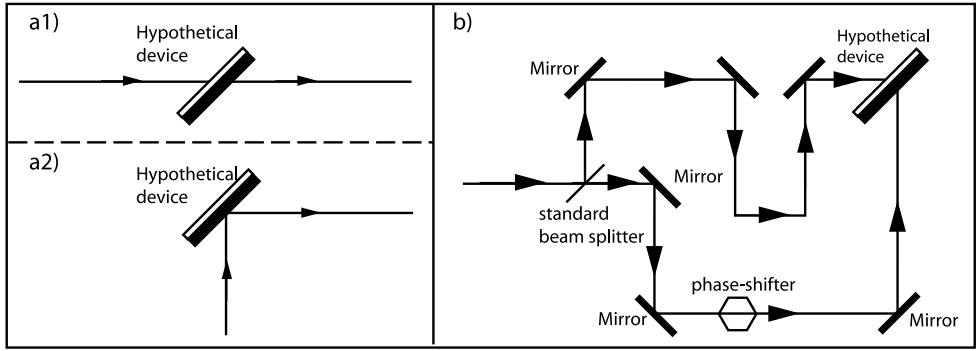}
\caption{The hypothetical device underlying the analysis of Ref.[1].
At left we have visualized its functioning for an horizontal a1) and
a vertical a2) particle, while at right we have illustrated how one
can resort to the device to transform a normalized wave function
into the neutral element of the Hilbert space, i.e. a function
vanishing everywhere, by making the paths of the two terms of the
superposition equal in length and by introducing the appropriate
phase-shifter  in such a way to make them interfere destructively. }
\label{f3}
\end{center}
\end{figure}

The final state $|\phi_{f}\rangle$ gives rise to a position density
distribution $|\phi_{f}^{\pm}(r)|^2 = |\langle
r\vert\phi_{f}^{\pm}\rangle|^2$ ($r$ being the running variable
along an axis perpendicular to the front of the incoming wave), like
those represented in Figs. 2.a) and 2.b). The superscript $+$ refers
to the fact that the arrangement is such to induce constructive
interference ($\varphi = 0$), while the superscript $-$ refers to
destructive interference ($\varphi = \pi$). We assume that the
detector is small enough to cover only the region right in front of
the two mirrors $C$.

In the case of constructive interference, the counter will fire with
probability $1$, since the final wave function $\phi_{f}^{+}(r)$ is
different from zero (practically) only in the interval of the
$r$-axis in which the counter is located, as made explicit in
Fig.2a). In the case in which in the lower arm of the interferometer
the phase shifter is introduced, at free will, then the two terms of
the superposition will interfere destructively in the (small) region
where the detector is placed along the $r$-axis. In this case the
counter will not fire (or better, the probability for it to fire is
negligibly small), since the final probability distribution for the
position of the particle is the one represented in Fig. 2.b), which
is associated to the wave function $\phi_{f}^{-}(r)$ .

As already remarked, the integral of the modulus square of the wave
function along the whole $r$-axis is, in either cases, always equal
to $1$. Given this fact, can we state, as the
author suggests, that the probability of the particle being found in
the region at right, due to the destructive interference,  is less
than $1$, and can even be made vanishingly
small? Obviously not: the whole process we have described is a
unitary evolution of the wave function crossing the interferometer, an evolution
which obviously preserves the norm of the state.

To summarize, the situation is the following. When the detector is
inserted at the output of the interferometer, with probability
$1$ it will fire in the case of
constructive interference (no phase-shifter inserted) and it will
(almost) surely detect nothing when the phase-shifter is inserted.
But does this mean that the norm of the initial triggering state has
been reduced? Not at all. This simply means that the particle is in
the region lying outside the one covered by the counter. Formally,
one can say that in this case the reduction has taken place to the
linear manifold associated to the projection operator $P_{out}$ on
the complement of the interval covered by the counter.

In brief,  a reduction process is actually induced by the presence
of the detector, but, for the two considered alternatives, the state
vector, even before the counter is inserted, is (practically) an
eigenstate either of the projection operator  $P_{in}$ on the
interval covered by the detector (if no phase-shift has been
introduced) or of the projection operator $P_{out}$ (if the
phase-shifter is present). Obviously, the effect of the reduction is
to transform the state vector $|\phi_{f}\rangle$ either in $P_{in}
|\phi_{f}\rangle$ or in $P_{out}
|\phi_{f}\rangle$, in the two
considered cases. Note that, in our case, only one of the norms  $ \| P_{in} |\phi_{f}\rangle \|$ and $  \| P_{out} |\phi_{f}\rangle \|$ is different from zero and the other equals 1.

\section{The correct version of Shiekh's proposal}

It is now an elementary task to make precise the situation devised
in Ref.~\cite{sh} and to show that no faster than light
communication can be achieved with the mechanism which the author is
proposing. Let us take once more the initial state vector considered
by the author:

\begin{equation}
\vert \psi,0\rangle= \frac{1}{\sqrt{2}}[\vert h+\rangle+\vert
h-\rangle];
\end{equation}
where both $|h+\rangle$ and $|h-\rangle$ are separately normalized.
Our analysis has shown that, in the case of constructive
interference, the final state will be:

\begin{equation} \label{eq:5}
\vert \psi,0\rangle\rightarrow\vert\psi_{f}^{+}\rangle=
\frac{1}{\sqrt{2}}[\vert \phi_{f}^{+}\rangle+\vert h-\rangle].
\end{equation}
The two terms in the square bracket are both normalized and
orthogonal (we neglect the overlapping of the well localized wave
functions), and the probability for the counter of the sender to
fire is $\langle \psi_{f}^{+} | P_{in} | \psi_{f}^{+} \rangle =
1/2$. Accordingly, the probability for the counter of the receiver
to fire is also $1/2$ since for a state like (5), one of the two
must fire.

When the phase-shifter is inserted, the analogous of the previous
equation is:

\begin{equation} \label{eq:6}
\vert  \psi,0\rangle\rightarrow \vert\psi_{f}^{-}
\rangle=\frac{1}{\sqrt{2}}[\vert \phi_{f}^{-}\rangle+\vert
h-\rangle].
\end{equation}
In this case, the counter of the sender will not fire for sure, but the
global state vector remains the equal weight superposition of Eq.(6) of two
normalized and orthogonal states. This means that the probability
for the counter of the receiver to fire is once more equal to 1/2: the
introduction of the phase-shifter has not influenced in any way
whatsoever this probability\footnote{Alternatively one might argue
as follows: let us replace the counter localizing the particle along
$r$ with three adjacent counters, one being the one already
considered and the other two covering the whole $r$ axis at left and
right of the first counter. Then one of the three counters will
detect the particle with probability 1/2, since the norm of each of the two states $\vert
\psi_{f}^{\pm}\rangle$ equals 1. Correspondingly reduction to a
state in the region of the counter or in the far-away region at left
will take place with equal probabilities.}.

\section{A clarifying example}

In this short section we want simply to reconsider the previous
argument reducing it to one in which the apparatus at right works
precisely as a Mach-Zender interferometer. As well known such an
apparatus works in the following way: if things are arranged to
yield constructive interference, then the impinging state is
transformed into a state propagating with certainty along the
horizontal direction, while if things are arranged to yield
destructive interference, then the impinging state is transformed
into a state propagating with certainty along the vertical
direction. Now, the appropriate way to discuss the situation is that
of inserting two detectors, one for horizontal and one for vertical
final directions of propagation, as we have shown in Fig. 4. Then,
if the norm of the impinging state $\vert \phi\rangle$ equals 1 and
constructive interference occurs, the detector along the horizontal
path will fire with certainty, while, for a destructive arrangement,
it is the one along the vertical path which will fire with
certainty.

Taking however into account that the initial state is the one of
Eq.~\eqref{eq:1}, by exactly the same argument of Section IV, we
conclude that the final state in this case would be the same as the
one of either Eq.~\eqref{eq:5} or Eq.~\eqref{eq:6}, with the
replacement of the states $\vert \phi_{f}^{+}\rangle$ and $\vert
\phi_{f}^{-}\rangle$ by two normalized states describing,
respectively, a particle propagating along the horizontal or one
propagating along the vertical direction. So, in accordance with the
fact that constructive or destructive interference is arranged by
the sender, either the detector along the horizontal or the one
along the vertical direction might fire, but, in both cases, the
probability that it will fire equals 1/2.  Correspondingly, the
probability of detecting the particle in the situation corresponding
to the evolved of the state $ \vert h-\rangle$, remains equal to 1/2
and cannot be changed at free will by the sender.
\begin{figure}[t]
\begin{center}
\includegraphics[scale=0.8]{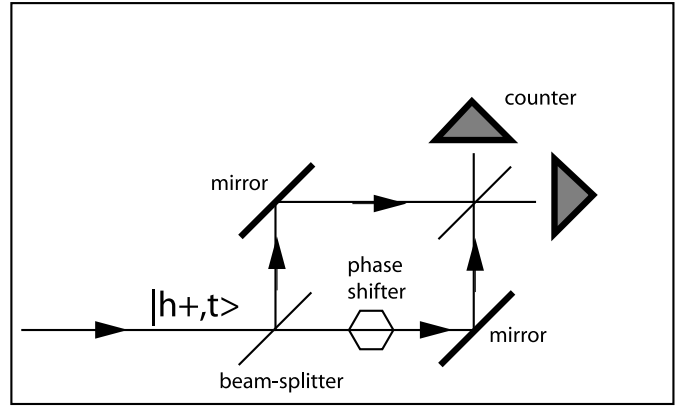}
\caption{The situation in the case in which the apparatus at right
is a standard Mach-Zender interferometer. In such a case only one of
the counters will fire, depending from the presence or absence of
the phase-shifter. The figure is more illuminating than the previous
ones because the occurrence of constructive and destructive
interference corresponds to the triggering of one or the other
detector and there is no need to consider the position distribution
along the whole $r$-axis.} \label{f4}
\end{center}
\end{figure}

\section{Conclusions}

We hope to have made clear that the proposal considered in
Ref.~\cite{sh} of resorting to wave packet reduction to send
superluminal signals does not work. This is not surprising: the
w.p.r. process, in spite of its basically nonlocal features, has
been conceived in such a way that it cannot lead to any
contradiction with relativistic requirements.

\section*{Acknowledgements}
The work of A.B. has been partly supported by DGF (Germany) and
partly by the EU grant ERG\;044941-STOCH-EQ.

\end{document}